\documentclass[aps,prd,preprint,groupedaddress,nofootinbib]{revtex4-1}

\pdfoutput=1
\usepackage{amsmath}
\usepackage{amssymb}
\usepackage{graphicx,mathrsfs}
\usepackage{nicefrac}

\newcommand{\be}{\begin{equation}}
\newcommand{\ee}{\end{equation}}
\newcommand{\bea}{\begin{eqnarray}}
\newcommand{\eea}{\end{eqnarray}}

\newcommand{\nn}{\nonumber}

\newcommand{\Y}{\mathcal Y}
\newcommand{\M}{\mathcal M}
 \newcommand{\LC}{\operatorname{LC}}
 \newcommand{\N}{\mathcal N}

\renewcommand{\sl}{/\!\!\!}

\begin{document}


\title{A Clockwork Solution to the Doublet-Triplet Splitting Problem}

\author{Gero von Gersdorff}

\affiliation{Departamento de F\'isica, Pontif\'icia Universidade Cat\'olica de Rio de Janeiro, Rio de Janeiro, Brazil}



\begin{abstract}

Maybe the biggest puzzle in grand unified theories (GUTs) is the apparent large splitting of the doublet and triplet Higgs masses. We suggest a novel mechanism to solve this puzzle, which relies on the clockwork mechanism to generate large hierarchies from order-one numbers. The tension between gauge coupling unification and proton lifetime from minimal $SU(5)$ GUTs is also removed in this scenario, and the theory remains perturbative until the Planck scale.

\end{abstract}


\maketitle

\section{Introduction}

The quantum numbers of the Standard Model (SM) fermions strongly suggest the existence of a unified gauge group such as $SU(5)$ or $SO(10)$. Moreover, the minimal supersymmetric (SUSY) extension of the SM, the MSSM,  
the gauge couplings unify with a good accuracy at a scale of the order of $\sim 10^{16}$ GeV \cite{Dimopoulos:1981yj,Ibanez:1981yh,Marciano:1981un,Amaldi:1991cn,Langacker:1991an,Ellis:1990wk}.
In the simplest $SU(5)$ scenario, the Higgs doublets unify with triplets in fundamental representations of the gauge group.
However, while the doublet Higgs bosons need to remain massless all the way down to the weak scale, their triplet partners need to be heavy in order to achieve coupling unification and suppress proton decay.
As we will review in the next section, such a splitting of doublets and triplets is highly unnatural in the minimal models, i.e., it requires an accurate cancellation of seemingly unrelated parameters.
Several extensions have been proposed to solve this doublet-triplet splitting problem (DTSP).
The elegant and minimal sliding singlet mechanism \cite{Witten:1981kv} turns out to fail under closer inspection, even though more complicated versions in $SU(6)$ extensions can work \cite{Sen:1984aq,Barr:1997pt,Maekawa:2003ka}.
Other solutions to the DTSP that have been suggested include the missing partner mechanism \cite{Masiero:1982fe,Grinstein:1982um,Hisano:1994fn}, pseudo Nambu-Goldstone bosons \cite{Inoue:1985cw,Anselm:1986um,Berezhiani:1989bd}, and extra dimensions \cite{Kawamura:2000ev,Kakizaki:2001en}.
For $SO(10)$, there is the possibility of the Dimopoulos-Wilczek mechanism \cite{Dimopoulos:1981xm}. 

In the $SU(5)$ context, the DTSP is actually a triad of three  interconnected problems which need to be dealt with together:
\begin{enumerate}
\item
The natural separation of doublet and triplet masses. Here, with natural we mean the absence of large cancellations between a priori unrelated free parameters \footnote{Since in SUSY theories the required operators appear in the superpotential, even in the presence of such tuning the theory is technically natural.}.
\item
Sufficient suppression of triplet-Higgs mediated dimension-five operators that trigger proton decay (PD).
\item
Precision gauge coupling unification (GCU): any additional gauge representation  at the GUT scale will contribute threshold corrections that change unification of gauge couplings.
\end{enumerate}
One could add to this a fourth problem, the so called $\mu/B\mu$ problem, that is, the question why the SUSY breaking soft mass between the two  Higgs doublets $B\mu$ is of the same order as the supersymmetric doublet mass $\mu^2$. A simple solution to the latter is found in the Giudice Masiero (GM) mechanism \cite{Giudice:1988yz}.

The idea we would like to propose to solve the above mentioned problems relies on the so-called clockwork mechanism \cite{Giudice:2016yja} which allows for generation of hierarchical couplings and scales from order-one numbers. 
The basic idea is to add copies of fields and employ spurious symmetries in order to enforce a special kind of mass matrix that only couples "nearest-neighbours" similar to a one-dimensional lattice Hamiltonian. Some of the light modes then exponentially localize at certain points in this lattice (or theory space), creating suppressed couplings with other low-energy modes. 
The clockwork mechanism has been used in many different contexts \cite{
Kaplan:2015fuy,Kehagias:2016kzt,Ahmed:2016viu,Hambye:2016qkf,Craig:2017cda,Teresi:2017yrp,Giudice:2017suc,Coy:2017yex,Ben-Dayan:2017rvr,Park:2017yrn,Lee:2017fin,Ibanez:2017vfl,Kim:2017mtc,Kehagias:2017grx,Ibarra:2017tju,vonGersdorff:2017iym,Patel:2017pct,
Choi:2017ncj,Giudice:2017fmj,Teresi:2018eai,Long:2018nsl,Kim:2018xsp,Niedermann:2018lhx,Agrawal:2018mkd,Alonso:2018bcg,Park:2018kst,Im:2018dum,Sannino:2019sch,Hong:2019bki,Kitabayashi:2019qvi,deSouza:2019wji,Folgado:2019gie,Kitabayashi:2020cpo,Bae:2020hys,vonGersdorff:2020ods,Kang:2020cxo,Babu:2020tnf,Joshipura:2020ibd} 
in order to create hierarchies in a natural way.
In the present implementation, due to a very mild accidental cancellation, the two physical Higgs doublets localize at opposite ends of the lattice, creating a small $\mu$ term, while  this localization does not take place for the remaining Higgs doublets and the triplets, leaving them to decouple at the GUT scale. We will also see that our mechanism allows to parametrically separate the effective triplet scales relevant for GCU and PD.

\section{Review of the problem(s)}

Here we briefly review the DTSP as it manifests itself in the minimal $SU(5)$ GUT model. The Higgs dependent terms in  the superpotential are
\be
W=\bar H(m_1+m_{24}Y)H+\Y_{ij} \bar HA_i \bar F_j+\frac{1}{2}\Y'_{ij} HA_iA_j
\label{eq:minimal}
\ee
The chiral superfields $\bar H$ and $\bar F_i$ transform in the $\bf \bar 5$ representation, $H$ in the $\bf 5$, and $A_i$ in the $\bf 10$.
Furthermore, $Y$ denotes hypercharge, 
\be
Y\equiv\begin{pmatrix}-\frac13\\&-\frac{1}{3}\\&&-\frac{1}{3}\\&&&\frac12\\&&&&\frac12\end{pmatrix}
\ee
and the coupling proportional to $Y$ in Eq.~(\ref{eq:minimal}) results from a trilinear $\bar H\Sigma H$ coupling of the Higgs fields with the $SU(5)$ breaking adjoint superfield $\Sigma$.
The DTSP arises from the fact that we need $\mu_D\equiv m_1+\frac{1}{2}m_{24}$ to be of the weak scale ($\mu_D$ is of course nothing but the $\mu$ parameter), while the corresponding parameter for the triplet, $\mu_T\equiv m_1-\frac{1}{3}m_{24}$ has to be of the order of the GUT scale. This is only possible if $m_1$ and $m_{24}$ are both of the order of the GUT scale, thus implying a delicate cancellation of the two contributions in $m_D$ at the level of one part in $10^{12}$. 

The reason why $\mu_T$ needs to be of the order of the GUT scale is twofold, which under closer inspection  reveals another problem. 
Firstly, let us consider gauge coupling unification (GCU). The weak scale gauge couplings $\alpha_i(m_Z)$ are a function of the unified high scale coupling $\alpha_5(\Lambda)$, and the masses of all the fields, at one loop they read \cite{Hisano:1992jj}
 \bea
 \frac{1}{\alpha_3(m_Z)}&=&  \frac{1}{\alpha_5(\Lambda)}+\frac{1}{2\pi}
 \left[
 -4\log\frac{m_{\rm susy}}{m_Z}
 -3\log\frac{\Lambda}{m_Z}
 \right.\nn\\
&& \left. 
 -4\log\frac{\Lambda}{m_V}+3\log\frac{\Lambda}{m_\Sigma}
 +\log\frac{\Lambda}{\mu_T}
 \right]\label{eq:beta3}\\
 \frac{1}{\alpha_2(m_Z)}&=&  \frac{1}{\alpha_5(\Lambda)}+\frac{1}{2\pi}
 \left[
  -\frac{25}{6}\log\frac{m_{\rm susy}}{m_Z}
 +\log\frac{\Lambda}{m_Z}
 \right.\nn\\
&& \left. 
 -6\log\frac{\Lambda}{m_V}+2\log\frac{\Lambda}{m_\Sigma}
 \right]\label{eq:beta2}\\
 \frac{1}{\alpha_1(m_Z)}&=&  \frac{1}{\alpha_5(\Lambda)}+\frac{1}{2\pi}
 \left[
  -\frac{5}{2}\log\frac{m_{\rm susy}}{m_Z}
 +\frac{33}{5}\log\frac{\Lambda}{m_Z}
 \right.\nn\\
&& \left. 
 -10\log\frac{\Lambda}{m_V}+\frac{2}{5}\log\frac{\Lambda}{\mu_T}
 \right]\label{eq:beta1}
 \eea
 where $m_V$ and $m_\Sigma$ are the $X/Y$ boson and adjoint scalar masses, and $\Lambda$ is any UV scale higher than the masses.
 For simplicity, we have considered a common sparticle mass $m_{\rm susy}$, our considerations will not depend on this assumption. 
 It is convenient to consider $\alpha_3^{-1}$ as well as the combinations $-2\alpha_3^{-1}-3\alpha_2^{-1}+5\alpha_1^{-1}$ and  $-2\alpha_3^{-1}+3\alpha_2^{-1}-\alpha_1^{-1}$.
 The former difference only depends on the combination $(m_\Sigma m_V^2)^\frac{1}{3}$ but is independent of  $\mu_T$ while the latter only depends on $\mu_T$:
\be
(-2\alpha_3^{-1}+3\alpha_2^{-1}-\alpha_1^{-1})(m_Z)
=
\frac{1}{2\pi}\left(\frac{12}{5}
\log\frac{\mu_T}{m_Z}
-2\log\frac{m_{SUSY}}{m_Z}\right)
\label{eq:betadiff}
\ee
This relation completely determines the triplet mass. For instance  Ref.~\cite{Murayama:2001ur}, including more realistic SUSY thresholds as well as some two loop corrections,   constrains $\mu_T$ to lie in the narrow corridor
\be
3.5\ 10^{14} {\ \rm GeV}<\mu_T<3.6\ 10^{15}{\ \rm GeV}
\label{eq:GCU}
\ee
at the 90\% confidence level.

It is well known that this value is in tension with the lifetime of the proton. In particular, integrating out the triplet Higgs gives rise to dimension-five  superpotential 
\be 
W_{d=5}=\frac{1}{\mu_T}\Y_{ij}\Y'_{kl}(Q_iL_j+\bar U_i\bar D_j)(Q_kQ_l+\bar U_k\bar E_l)
\label{eq:dim5}
\ee   
The $QQQL$ and $\bar U\bar U\bar D\bar E$ operators violate Baryon number leading to PD via loops, which induce a bound on $\mu_T$ \cite{Murayama:2001ur}
\be
\mu_T>7.6\ 10^{16}\ {\rm GeV} 
\label{eq:PD}
\ee
that is evidently in conflict with Eq.~(\ref{eq:GCU}).
These bounds apply to generic sfermion masses and mixings. It has been pointed out that decoupling the first two sfermion generations and choosing a peculiar pattern of 
sfermion mixings with the third generation, one can tune the proton decay constraints away \cite{Bajc:2002bv}.
Another option would be to push the scale of all sfermion masses up. However, this requires giving up on naturalness more than presently required experimentally.
For the present paper, we are assuming a generic sfermion spectrum with the benchmark bounds Eq.~(\ref{eq:GCU}) and (\ref{eq:PD}), and explore how our DTS mechanism can ease the tension between GCU and PD. 

\section{Model}

The basic idea behind our proposal is to clone the Higgs sector to include $N$ Higgs fields of $\bf 5 $ and $\bf\bar 5$ each, with a clockwork-type mass matrix. 
The superpotential is taken to be 
\be
W=\bar H^T (\M_{1}+\M_{24}Y)H+\Y_{ij} \bar H_1A_i \bar F_j+\frac{1}{2}\Y'_{ij} H_NA_iA_j.
\ee
The mass matrices $\M_1$ and $\M_{24}$ are taken of the following structure
\be
\M_1=\alpha_1 M-\beta_{1}K\qquad \M_{24}=\alpha_{24}M-\beta_{24}K
\label{eq:alphabeta}
\ee
where $\alpha_{1,24}$ and $\beta_{1,24}$ are dimensionless constants, 
and the  the $N\times N$ matrices $M$ and $K$ are two spurions   given by 
\be
(M)_{ij}=m_i\delta_{ij}\qquad (K)_{ij}=k_i\delta_{i,j+1}
\ee
that is, the mass matrix has all  $m'$s in the $N$ diagonal entries and $k'$s in the $N-1$ lower sub-diagonal ones.
 In the absence of the superpotential, $W=0$, the theory has a large
$G=U(N)'\times U(N)$ chiral symmetry
 acting on $\bar H$ and $ H$ respectively (we take $H$ transforming in $( 1,N)$ and $\bar H$ transforming in $ (\bar N,1)$ ). The spurions $M$ and $K$ then transform in $(N,\bar N)$.
The stabilizer groups  are  the Abelian subgroups $H_M=U(1)^N$ and $H_K=U(1)^{N+1}$ generated by
\be
\mathfrak h_M=\{Q'_i+Q_i\}
\qquad
\mathfrak h_K=\{Q'_1,Q_N,Q'_{i}+Q_{i-1}, i>1\}
\ee 
The intersection of the two is  the vectorlike $U(1)$ generated by $\sum_i Q_i+Q'_i$.
The Yukawa couplings
 $\Y$ and $\Y'$ transform in the $(N,1)$ and $(1,\bar N)$ respectively and  
  leave the groups
$H_\Y=U(N-1)'\times U(N)$ and 
$H_{\Y'}=U'(N)\times U(N-1)$ unbroken, which breaks the remaining $U(1)$.
We will assume these are the only spurions breaking $G$, and explore the phenomenology of this ansatz.

Let us then define  the doublet and triplet  mass matrices ($X=D,T$)
\be
\M_X\equiv M_X-K_X
=\begin{pmatrix}
m_{X,1}&&&\\
-k_{X,1}&m_{X,2}&&&\\
&-k_{X,2}&m_{X,3}&&\\
&&\ddots&\ddots\\
&&&-k_{X,N-1}&m_{X,N}\\
\end{pmatrix}
\ee
where 
\be m_{D,i}\equiv \left(\alpha_1+\frac{1}{2}\alpha_{24}\right)m_i\,,\qquad k_{D,i}\equiv \left(\beta_1+\frac{1}{2}\beta_{24}\right)k_i\,,
\ee
 and  
\be m_{T,i}\equiv \left(\alpha_1-\frac{1}{3}\alpha_{24}\right)m_i\,,\qquad k_{T,i}\equiv \left(\beta_1-\frac{1}{3}\beta_{24}\right)k_i\,.
\ee
The mass matrices $\M_{D,T}$  are of the clockwork type \cite{Giudice:2016yja}, however with the difference that they have equal number of rows and columns (the spectrum is vector-like), and hence there is no chiral zero mode.
What is peculiar in this vector-like clockwork is that for $m_{D,i}\ll k_{D,j},$ there is a single mode with an exponentially suppressed  mass
\be
\mu_{D,1}\approx \frac{\prod_{i=1}^N m_{D,i}}{ \prod_{i=1}^{N-1}k_{D_i}}
\label{eq:light}
\ee
 with all the other modes of order $\mu_{D,i+1}\approx k_{D,i}$ $(1\leq i<N)$. 
 This result can easily be obtained by treating  $M_D$ as a small perturbation to $K_D$ and integrating out the vectorlike pairs with masses $k_{D,i}$.
 Notice that setting any of the $m_{D,i}$ to zero restores a chiral $U(1)$ and thus creates a pair of zero modes, implying that all of the $m_D$ must be nonzero to avoid this case (this is also clear from looking at the determinant fo $M_D$). 
 If, say, $m_\ell$ vanishes we end up with two traditional chiral clockwork chains, with (for $m_{D,i}\ll k_{D,j}$) one $ \bar H $ zero mode localized near site $0$ and one  $H$ zero mode localized near site $N$. Their wave functions are suppressed at site $\ell$, and turning on $m_\ell$ gives thus an exponentially suppressed Dirac mass to this pair of chiral zero modes. In fact, 
 since there is a large mass gap between the lightest and the other mass eigenvalues, we can use the prescription of Ref.~\cite{vonGersdorff:2019gle} to calculate the corresponding eigenvectors. To this end, define a function $\LC$ which when acting on a matrix selects its longest column (in the canonical norm of vectors).
For $\N\equiv \M_D^{-1}$, the infinite sequence of vectors
 \be
\LC(\N)\,,\ \LC(\N \N^\dagger)\,,\ \LC(\N\N^\dagger \N)\,, ...
 \ee
 rapidly converges to the (unnormalized) eigenvector of $\M_D^\dagger\M_D$ to eigenvalue $|\mu_{D,1}|^2$ \cite{vonGersdorff:2019gle}.  For the corresponding eigenvector of $\M_D\M_D^\dagger$, one substitutes $\N\to\N^\dagger$. For instance, for $m_{D,i}=m_D$ and $k_{D,i}=k_D$ this gives the leading approximations for the wave functions
 \be
 f_i\approx \left(\frac{m_D}{k_D}\right)^{N-i}
 \qquad  \bar f_i\approx \left(\frac{m^*_D}{k^*_D}\right)^{i-1}
 \ee
 such that the lightest fields can be written approximately as $h=f^\dagger H\approx H_N$ and $\bar h=\bar f^\dagger \bar H\approx \bar H_1$ respectively. 
 We see  that the two chiralities participating in the formation of the $\mu$ term localize at opposite ends of the theory space.
The characteristics of this light mode is reminiscent of the lightest fermion mode in five dimensional compactifications on the interval $[0,L]$ with twisted boundary conditions $(1+\gamma_5)\psi(0)=0=(1-\gamma_5)\psi(L)$.


 We will now suppose a partial cancellation to happen between the two parameters $\alpha_1$ and $\frac{1}{2}\alpha_{24}$, such  that $m_{D,i}\ll k_{D,i}$ (we will quantify this in  a moment),  and a light mode given by Eq.~(\ref{eq:light}) exists. Of course, such a cancellation cannot occur simultaneously for the triplets, and such a light mode is absent.
 In the following, we will take all the $m_i=m$ and $k_i=k$, for unequal parameters we need to replace $k^{N-1}$ and $m^N$ by the products of the individual parameters in all of the expressions below, this will not change our conclusions. 
 The $\mu$ term is then simply given as
\be
\mu\equiv \mu_{D,1}\approx \frac{(m_D)^N}{(k_D)^{N-1}}
\ee
 Moreover, we can derive very simple exact formulas regarding GCU and dimension-five PD. For GCU, each additional $5+\bar 5$ with masses $\mu_T'$ and $\mu_D'$ contribute $\frac{1}{2\pi}\Delta_i$ to the gauge couplings in Eq.~(\ref{eq:beta3})--(\ref{eq:beta1}) where
 \be
\Delta_3=\log\frac{\Lambda}{\mu_T'}\,,\qquad 
\Delta_2=\log\frac{\Lambda}{\mu_D'}\,,
\ee
\be
\Delta_1=\frac{3}{5}\log\frac{\Lambda}{\mu_D'}+\frac{2}{5}\log\frac{\Lambda}{\mu_T'}\,.
\ee
Then, in Eq.~(\ref{eq:betadiff}) we must replace $\mu_T$ with an effective triplet mass
\be
\mu_T\to \mu_T^{\rm eff}
\equiv \left|\frac{\mu_{T,1}\cdots\mu_{T,N}}{\mu_{D,2}\cdots\mu_{D,N}}\right|
=\frac{|\det \M_T|}{|\det \M_D|}|\mu|\approx\frac{|m_T|^N}{|k_D|^{N-1}}\,.
\ee
It is this quantity to which the constraint Eq.~(\ref{eq:GCU}) applies.

On the other hand the coefficient of the dimension-five operator in Eq.~(\ref{eq:dim5}) is replaced by
\be
\frac{1}{\mu_T}\to
\frac{1}{\tilde \mu^{\rm eff}_T}\equiv (\M_T^{-1})_{N1}=\frac{(k_T)^{N-1}}{(m_T)^N}\,.
\ee
Now, the constraint in Eq.~(\ref{eq:PD}) applies to $\tilde \mu_T^{\rm eff}$, which is different from $\mu_T^{\rm eff}$.
More explicitly, we have
\be
\left|\frac{\tilde \mu_T^{\rm eff}} {\mu_T^{\rm eff}}\right|=\left|\frac{k_D}{k_T}\right|^{N-1}\!\!\!.
\ee
For instance, consider the case where $\beta_1=0$ in Eq.~(\ref{eq:alphabeta}), then $k_D/k_T=\frac{3}{2}$ and for small to moderate $N$, one very efficiently separates the two scales and completely removes the tension of the minimal $SU(5)$ model.
As far as this separation is concerned, our mechanism is related to the idea of Ref.~\cite{Babu:1993we} which considers a similar mass matrix with $N=2$. 

Next, consider the ratio
\be
\left|\frac{\mu_T^{\rm eff}}{\mu}\right|=\left|\frac{m_T}{m_D}\right|^N
\label{eq:hierarchy}
\ee
For a $|\mu|\sim 1$ TeV, we find that we need $|m_T/m_D|\approx 6.3$ (15.8) for $N=15$ ($N=10$). The cancellation between $\alpha_1$ and $\alpha_{24}$ for this hierarchy is at the level of 25\%  (10\%), in other words  mild when compared to the original doublet-triplet splitting.

\begin{figure}
\includegraphics[width=7 cm]{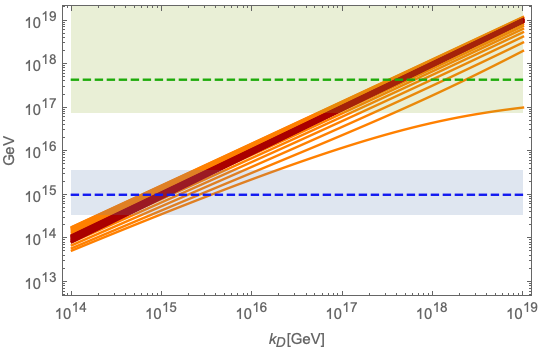}
\caption{Spectrum for the heavy doublets (red) and triplets (orange) as a function of the parameter $k_D$, for $\mu=1$ TeV and $N=15$. 
The blue and green dashed lines are the parameters $\mu_T^{\rm eff}$ (fixed to $10^{15}$ GeV) and $\tilde\mu_T^{\rm eff}$ respectively, and the shaded regions mark their allowed values (from GCU and PD respectively).}
\label{fig:spec}
\end{figure}

Notice that the overall scale of the mass parameters $k$ and $m$ is not fixed by these considerations alone. Setting $\beta_1=0$, we are left with three more parameters $\beta_{24} k$, $\alpha_{1}m$ and $\alpha_{24}m$, two of which can be eliminated by fixing $\mu$ and $\mu_T^{\rm eff}$. We take as the remaining free parameter $k_D=\frac{1}{2}\beta_{24}k$, which is to a good approximation the (nearly degenerate) mass of heavy  doublets. We show in Fig.~\ref{fig:spec} the spectrum as a function of $|k_D|$. 
For increasing $k_D$, the GUT breaking vacuum expectation value  $v_{24}$, defined by $\langle\Sigma\rangle=v_{24}Y$ should also go up to avoid a nonperturbative $\bar H \Sigma H$ coupling. This means that in this region, the dimension-six proton decay operators are more suppressed, while there is a  splitting  $m_V\gg m_\Sigma$  as $m_\Sigma m_V^2$ is also tightly constrained by GCU.

Since we are adding a potentially large number of representations, one might wonder if the gauge coupling becomes non-perturbative below the Planck scale. However, this only happens when $N\gtrsim 40$, and such large numbers are not required to solve the DTSP. In this sense there is an advantage over the missing partner mechanism, which employs exotic representations with large Dynkin indices. For instance, the ${\bf 50+\overline{ 50}+75 }$ \cite{Masiero:1982fe} contribute to the running as much as the equivalent of 60 ${\bf 5+\bar 5}$ representations, leading to the well-known non-perturbativity issues of that model that need to be resolved with more sophisticated model building \cite{Hisano:1994fn}.

Let us finally comment on the $\mu/B\mu$ problem. A simple way to implement the GM mechanism is to demand that the doublet mass $\mu_{D,1}$ created from the present mechanism is actually even smaller than the TeV scale and that there is another (dominant) contribution to $\mu$ coming from the operator $X^*\bar HH$ in the K\"ahler potential (whereas the  $B\mu$ term is coming from $|X|^2 \bar HH$). These contributions are of course completely negligible for all the other doublets and triplets.

In summary, we have presented a novel mechanism to solve the DTSP in $SU(5)$ grand unification models, which at the same time  removes the tension between the limit on the triplet scales for GCU and PD.

\begin{acknowledgments}
This work was financed in part by the Coordena\c c\~ao de Aperfei\c coamento de Pessoal
de N\'ivel Superior - Brasil (CAPES) - Finance Code 001, and by the Conselho Nacional de Desenvolvimento Cient\'ifico (CNPq) under grant number 422227/2018-8.
I would like to thank Arman Esmaili for feedback on the manuscript.
\end{acknowledgments}

\bibliography{paper,clockworkpapers}

\end{document}